\documentclass[12pt,epsf,psfig]{article}
\usepackage{subfigure}
\usepackage{psfig}

\def\reference{\parskip 0pt\par\noindent\hangindent 0.5 truecm}

\def\Ha {H$\alpha$}
\def\Hb {H$\beta$}
\def\etal {{\it\ et al}}
\def\deg {{$^o$}}
\def\eg {{\it e.g.}}
\def\ie {{\it i.e.}}
\def\kms {km s$^{-1}$}
\def\arcsec {$^{\prime\prime}$}

\begin{document}
\title{\bf Taurus Tunable Filter -- seven years of observing}
\author{J. Bland-Hawthorn $^1$ \and  L. Kedziora-Chudczer $^2$
}

\date{}
\maketitle

{\center $^1$ Anglo-Australian Observatory, PO Box 296, Epping, NSW 2121\\jbh@aao.gov.au\\[3mm]
$^2$ Dept. of Astrophysics \& Optics, School of Physics, University of New South Wales, NSW 2052\\lkedzior@physics.usyd.edu.au\\[3mm]}

\begin{abstract}
The Taurus Tunable Filter (TTF) has now been in regular use for seven
years on the Anglo-Australian Telescope (AAT). The instrument was also
used for three years (1996--1999) on the William Herschel Telescope
(WHT).  We present a brief review of the different applications in
order to illustrate the versatility of tunable filters in optical/IR
spectrophotometric imaging.  Tunable filters are now planned or under
development for 6-10m class telescopes which ensures their use for
years to come.
\end{abstract}

\section{INTRODUCTION}

We provide a short review of the first seven years of observing with
the Taurus Tunable Filter at the AAT; for the first three years,
the TTF was also used at the WHT. The instrument has had most
of its use on the AAT where it gets between 10--15\% of the telescope
time scheduled by the PATT (UK) and ATAC (Australia) committees.  In a
nutshell, the TTF allows for wide-field (10$^{\prime}$) spectrophotometric
imaging from 370nm to 1000nm with resolving powers generally in the range
100 to 1000. An important feature of the instrument is the use of charge
shuffling synchronized to band switching in order to greatly suppress
systematic errors associated with conventional imaging. Another aspect of
the TTF is time series readout coupled to band switching which has lead
to important new work on compact variable sources.  Technical accounts
of the instrument, and its related charge shuffling modes, can be found
in Appendices A and B.

There have been several recent reviews on tunable filters
including Bland-Hawthorn (2000$a,b$), Jones (2001$a$) and 
Jones\etal\ (2001).
Recent scientific reviews can be found in Veilleux\etal\ (2002)
and Veilleux (2002$a,b$). A list of largely unexplored science
areas with tunable filters is given in Bland-Hawthorn\etal\ (2001).

The purpose of this review is to stress the versatility of tunable
filters in observational astronomy.  The Taurus focal reducer
has also produced important results in its other modes of operation
(see Appendix A):
polarimetry, Fabry-Perot `staring' and emission line scanning, and most
recently multi-object spectroscopy (e.g. Riddick, Lucas \& Roche 2003).
But here we concentrate specifically on the TTF mode of operation.

All observations discussed here were undertaken on 4m telescopes, often
in non-ideal observing conditions. These authors eagerly anticipate the
tunable filters which are planned or under way for the new generation
optical/IR telescopes on superb sites. These include, inter alia, the
Osiris tunable filter on the Grantecan 10.2m telescope
(Cepa\etal\ 2002), a possible tunable filter within FORS on the VLT,
the proposed Maryland Magellan tunable filter (MMTF), and the tunable
filter within the Goodman spectrograph on the SOAR 4m telescope. When
one reviews the results below, it is important to keep in mind that the
new generation of tunable filters need only improve on one of the
following -- site conditions, instrument performance, field of view,
pixel sampling, telescope aperture -- to achieve a major gain.  
\begin{figure}
\psfig{bbllx=17pt,bblly=146pt,bburx=592pt,bbury=718pt,file=tresse.ps,angle=0,height=95mm,width=95mm}
\caption{
The evolution of the H$\alpha$ luminosity function with redshift 
taken from Tresse\etal\ (2001; Fig. 13). The filled circles are the 
data points from their ISAAC/VLT survey.
The short-long dashed curves are the preliminary H$\alpha$ LF
from Jones \& Bland-Hawthorn (2001) from left to right
respectively at $z=0.08$, $z=0.24$, $z=0.40$. The other curves are
defined in Tresse\etal\ (2001).
}
\label{fig1}
\end{figure}

\section{SURVEYS OF STAR FORMING GALAXIES}

\subsection{Field Galaxies}

Tunable filters are ideally suited to surveys
of star forming galaxies in different environments. The object selection
is based on the property we are trying to measure, \ie\ the
star formation rate via the \Ha\ line. This was the principle
behind the TTF Field Galaxy Survey which was the basis of D.H. Jones'
thesis (2001). He obtained photometric \Ha\ data and restricted \Hb\
data on 4 clusters and 9 field positions.  All observations were highly
successful in identifying line emitting galaxies, typically $10-40$
objects above $3\sigma$ per pointing (although rather fewer in \Hb),
finding many more objects than comparable studies with conventional
imaging techniques (\eg\ Hu \& McMahon 1996; Thommes\etal\ 1998).  
The first part of TTF Field Galaxy Survey was
published in Jones \& Bland-Hawthorn (2001).

Recent evidence suggests a decline in the volume-averaged star
formation rate (SFR) with the advance of cosmic time since $z\sim 1$.
The survey set out to derive \Ha\ luminosity functions in three
discrete wavelength intervals at $z=0.08$, 0.24, and 0.39. One of the
interesting surprises was the discovery of a population of compact
sources found to have moderate amounts of \Ha\ emission: some of these
sources would have been dismissed as stars in earlier photographically
selected surveys.

The TTF Field Galaxy Survey measures volume-averaged star formation
rates at intermediate redshift that are a factor of $2-3$ times higher
than those measured from multi-slit or multi-fibre surveys. Since these
results have been largely confirmed by the CADIS emission line survey
(Hippelein et al 2003) and the SUBARU emission line survey (Fujita et
al 2003), it is likely that the restricted apertures of fibres and
slits have underestimated the total H$\alpha$ line flux at low
redshift.

In a new study of galaxies in the Canada-France Redshift Survey (CFRS),
Tresse\etal\ (2002) combine the TTF Field Galaxy Survey measurements
with their ISAAC/VLT \Ha\ luminosity function at $z\sim 1$ (see
Fig.~1).  They find that the comoving \Ha\ luminosity density
increases by a factor of 12 from $z=0.2$ to $z=1.3$. Their results confirm
a strong rise of the star formation rate at $z<1.3$ proportional to
$(1+z)^{4.1\pm 0.3}$.

Tober\etal\ (2003; see also Glazebrook 1997) carried out a deep survey
of 20 square arcminutes in the Hubble Deep Field North (HDF-N) using the 
Taurus Tunable Filter and William Herschel Telescope (WHT) to observe in
contiguous sequences of narrow band slices in the
710$-$910 nm region of the spectrum.  These authors sampled almost
70 narrowband images in four redshift intervals.  In this way, one gets
a {\it multiplicity effect} in that the same images can be used to find
different sources at different redshifts, thereby allowing for a
derivation of the star formation rate over a wide range in redshift.
The data, which were taken during four long dark nights at the WHT,
have excellent image quality with psf $\approx$ 0.7\arcsec\ FWHM after
combining the full data set.

Cross-matching with Keck spectroscopy of the HDF-N results in
a line-luminosity limited sample, with very highly complete redshift
identification, containing OII, \Hb\ and \Ha\ emitters over the $z =
0.3 - 0.9$. From this direct star-formation rate selected sample they
reconstruct, via maximum likelihood techniques, the star-formation
history of the Universe to $z=1$. Their study finds no evidence for any
new population of low-luminosity strong-lined objects to an approximate
flux limit of $2\times 10^{-20}$ W m$^{-2}$ and covering a cosmological
volume of 1000 Mpc$^3$ at $z=1$.  From their complete narrow-band
sample of a single sight line, Tober\etal\ find a star-formation history 
consistent with earlier estimates from broad-band selected samples.

\begin{figure}
\vskip 2.7in
\includegraphics{jones1.ps}
\includegraphics{smail.ps}
\caption{
{\sl Left --}
Distribution of emission-line objects from a $z=0.4$ cluster
field if all of the emission is assumed to be redshifted H$\beta$.
The inset shows the TTF passbands relative to the wavelength of
redshifted H$\beta$ (dotted line). {\sl Right --}  Illustration of
where the line emitters fall in a cluster colour magnitude diagram.
The cluster data are from Smail\etal\ (1998).
}
\label{fig2}
\end{figure}

\begin{figure}
\vskip 2.7in
\includegraphics{jones3.ps}
\includegraphics{jones2.ps}
\caption{
TTF spectra (Left) and scans (Right) for a sample of emission-line
cluster candidates. The number on the left of each panel is the object
ID, the numbers at right are the flux in counts (top) and shape
classification parameter (bottom; 1=stellar, 0=galaxian).  No attempt
has been made to reject possible fore/background contaminants on the
basis of appearance or redshift. The vertical dotted line denotes the
cluster redshift from Keck spectra of $\sim 10$ galaxies in the field.
The wavelength shift of spectra between objects is due to the off-axis
phase effect across the TTF field (Bland-Hawthorn \& Jones 1998).
}
\label{fig3} 
\end{figure}

\subsection{Clusters}

Until now, identifying the majority of cluster members from a
combination of kinematics and a colour-magnitude diagram has been
difficult, particularly for blue objects, due to contamination of
sightline galaxies. But TTF observations can be used to identify faint
cluster members unambiguously, in particular those with line emission.
These objects are typically blue so the TTF is an ideally suited to
unscrambling the region of contamination (see Fig.~2({\em Right})).  We
find that k$+$a galaxies (Dressler \& Gunn 1983) can also be identified
from \Ha\ absorption (Jones \& Bland-Hawthorn 1999, Fig.~1).

In Fig.~3, we demonstrate the power of TTF to find line
emitters even in poor seeing (2$^{\prime\prime}$). Dalcanton (1996) and
Zaritsky\etal\ (1997) use CCD drift scanned images to identify clusters
through enhanced surface brightness fluctuations out to $z\approx
1.1$.  For one cluster at $z=0.45$, in just 2 hours, we identified 10
cluster members at the AAT in 2\arcsec\ seeing, compared to 6 members
in 1.5 nights using the Keck LRIS spectrograph.

For the rich cluster A3665 (AC 106), first studied by Couch \& Newell
(1984), Zaritsky \& Jones identify 40 \Ha\ emitting candidates above 
3$\sigma$, with many more candidates at lower thresholds (down to 0.03
solar masses per year at $z=0.25$), in just four hours.  Example TTF
scans are shown in Fig.~3({\em Right}), with the \Ha\ line profiles in
Fig.~3({\em Left}).

To date, average star formation rates in $z=0.2-0.5$ clusters
appear to be quite low compared to the field (cf. Couch\etal\ 2001).

\subsection{Quasar environment}

Baker\etal\ (2001) used the [OII]372.7nm line to detect starforming
galaxies in the vicinity of MRC B0450-221. Many emission line galaxies
can be visible over the whole field (down to a few
solar masses per year). Their projected separations from the quasar
range from 200 to 700 kpc. A nice demonstration of the power of
TTF imaging is the detection of very faint extended [OII] 
emission along the radio axis of the quasar.

\subsection{Quasar sight lines}

Francis\etal\ (2001$a$,$b$) have continued their study of the `Francis
cluster', an extended ensemble of galaxies at $z=2.38$ that was
originally identified from the damped Ly$\alpha$ absorption occurring
at the same redshift in a paired quasar sightline. In this study, they
look for Ly$\alpha$ emission from star forming galaxies within the
ensemble.  It is clear that the future holds great promise for using
quasars to identify gas-rich starforming environments at high
redshift.

\section{ENERGETIC GAS IN CLUSTERS}

There have been claims of EUV excesses in clusters which, if
true, would suggest that 10$^{5-6}$K gas accounts for a large fraction
of missing baryons (Lieu\etal\ 1999; Bowyer\etal\ 1999).  EUV emission
is exceedingly difficult to detect at Earth:  it is too hot to be
detected optically, too cool to be seen by x-ray satellites, and
undergoes molecular line absorption as it propagates through the ISM.

Using the EUVE satellite, Lieu\etal\ (1999) claim a direct EUV
detection  of the cluster Abell 1795.  Maloney \& Bland-Hawthorn (2001)
found that such a strong EUV field would ionize the molecular and HI
disks of all spirals in the cluster to such an extent that they should
all be \Ha\ bright.  TTF observations of the cluster were kindly taken
by A. Edge in `straddle shuffle' (Maloney \& Bland-Hawthorn 2001) mode in order to subtract the
continuum light of the cluster very accurately. Only very faint
\Ha\ emission was found in the cluster spirals.

Jaffe \& Bremer (2000) used `on-off' charge shuffling in order to
detect very faint levels of \Ha\ emission in cooling flow clusters.  In
the case of Abell 2597, they find a nebula extending over 50$-$60 kpc
from the cluster centre. A. Edge and collaborators have looked at a
large sample of cooling flow clusters and detect optical line emission
in all cases.  

\section{GRAVITATION LENSING}

Tunable filters are ideally suited to observe extended sources, in particular,
irregularly shaped sources with no axis of symmetry. Earlier work with
D.H. Jones on star formation in clusters occasionally turned up
spectacular gravitational lenses throughout the cluster.  As seen
through the TTF, these sources can have high contrast because the
background object is commonly a high-redshift starforming galaxy which
emits strongly in emission lines.  Hewett\etal\ (2000) used the TTF
to detect extended emission in a galaxy-galaxy gravitational lens due
to an intermediate redshift elliptical lensing a $z=3.59$ star forming
galaxy.  They were following up a suggestion by Miralde-Escud\'{e} \&
Lehar (1992) that the surface density of high-redshift starforming
galaxies is so high, galaxy-galaxy lensing should be relatively
common.

\section{GALAXIES}

The tunable filter has the major advantage of providing
detailed spectrophotometric information over a larger field of view
(FOV) than that of other 3D instruments. It would be very difficult
\begin{figure}
\psfig{bbllx=51pt,bblly=34pt,bburx=539pt,bbury=526pt,file=tadhunter.ps,angle=0,height=85mm,width=85mm}
\caption{Deep TTF image of Coma A in \Ha\ from Tadhunter\etal\ (2000). A complex
spiral nebula is seen to extend over 150~kpc. The overlaid radio 
continuum map (contours) shows striking similarities to the distribution
of ionized gas.}
\label{fig4}
\end{figure}
and very expensive to build an integral field spectrograph which
could provide the same field and the same photometric integrity at
low light levels. The tunable filter is therefore ideally suited 
to study nearby galaxies where the line-emitting gas extends over several 
arcminutes. This ionized material is an excellent probe of the phenomena 
taking place in the core of starburst and active galaxies, and can be 
used to quantify the impact of nuclear and star-formation activity on the
environment and vice-versa. Once again, the reader is encouraged to
look at recent reviews by Veilleux (2002$a,b$).

\subsection{Radio galaxies}

Tadhunter and collaborators have undertaken a comprehensive survey of
extended ionized haloes around powerful radio galaxies (Tadhunter\etal\
2000; Solorzano-Inarrea\etal\ 2002; Solorzano-Inarrea \& Tadhunter 2003).  
It is well known that optical
line emission often aligns with the radio axis, as these authors find
in all cases. However more surprising was the frequency of optical
emission perpendicular to the radio axis. In Fig.~4, the optical
emission in Coma A extends over more than 150 kpc in an egg-shaped nebula
(Tadhunter\etal\ 2000). More remarkably, the faint outer nebula
emission matches the extended outer radio lobe over these same scales.

\begin{figure}
\psfig{bbllx=-110pt,bblly=48pt,bburx=721pt,bbury=740pt,file=shopbell.ps,angle=0,height=90mm,width=90mm}
\caption{
The extended ionized nebula surrounding the x-ray selected quasar
MR~2251-178.  The top figures show deep TTF \Ha\ images in two closely 
spaced narrow bands in order to emphasize the kinematic structure.
The summed emission (bottom right) is seen to extend over 200~kpc or 
more. The stellar continuum image is also shown (bottom left).
}
\label{fig5}
\end{figure}

The frequent match-up between optical, radio and x-ray emission in
radio galaxies, Seyfert and starburst galaxies is a topic of great
interest (see below). In the case of low power radio jets, the emission is thought to
arise from sychrotron, whereas in high power radio jets, the x-rays
appear to arise from inverse Compton emission (Birkinshaw\etal\ 2001).
\subsection{Quasar nebulae}

Shopbell\etal\ (1999) recently published a remarkable extended nebula
surrounding the x-ray selected quasar, MR~2251-178.  The summed image
in Fig.~5 is one of the faintest extended emission-line images
ever published. The nebula, which extends over more than 200~kpc, shows
pronounced tidal arm structure. These authors chose two narrow bands
spaced by half a bandwidth in order to study the kinematics. The entire
structure appears to be in rotation about the quasar. The gas is not
obviously connected with any other companions in the field.  UV
radiation from the quasar would have to be escaping almost
isotropically to account for the ionization.  Both the source of the
gas and the source of ionization remain a mystery.

\subsection{Seyfert \& starburst galaxies}

Seyfert galaxies often show the impact of the nuclear activity on the
very large scale disk of a spatially resolved galaxy in the near field.
Two of the most recent and spectacular examples are NGC 1068 and
NGC 7213. Can we use the ISM as a screen to interpret what is essentially
unresolved and unseen at the core?

NGC 1068 is one of the most remarkable of Seyfert galaxies in the near
field. It is well known that the core harbours a concealed Seyfert 1
(`low power quasar') nucleus, but just how energetic is the source? 
The radio jet axis exhibits a wide range of activity (Cecil\etal\ 2002)
but can we infer the true energetics and nuclear spectrum from 
multi-wavelength studies (Alexander\etal\ 2000; Pier \& Krolik 1993)?

Shopbell\etal\ (2001; see also Veilleux\etal\ 2001) find that the
well known ionization cone (Pogge 1988) extends up to and {\it beyond}
the HI edge of the early-type spiral. The observed nebula requires
rather special conditions to be detectable over such a huge extent.
The most likely interpretation is that a multi-phase, vertically
distended ISM is being blasted by a very energetic central source
($L_{uvx} > 5\times 10^{43}$ erg s$^{-1}$).  Indeed, the Chandra
observations appear to show the most spectacular example to date of an
`x-ray ionization cone' (Young\etal\ 2001).

These galaxy-scale ionization cones are not unique to NGC 1068. Another
has been detected with the TTF in NGC 7213 (Rupke\etal\ 2002; Veilleux\etal\
2001) to rival the spectacular system in NGC 5252 studied by Tadhunter
\& Tsvetanov (1989).  Ionization cones are often associated with
nuclear jets.  Recent examples from TTF observations include IC 5063
(Cecil\etal\ 2001) and Circinus (Veilleux \& Bland-Hawthorn 1997).
Circinus, a large spiral close to the Galactic plane, is particularly
noteworthy: the system shows evidence for a whole network of `artillery
shells' blasting away from the nucleus. One such filament shows a
spectacular `Herbig Haro'-like morphology.

\begin{figure}
\centering
\mbox{\subfigure[]{\psfig{bbllx=84pt,bblly=190pt,bburx=527pt,bbury=602pt,file=hawthorn1.ps,angle=0,height=70mm,width=70mm}}\quad
\subfigure[]{\psfig{bbllx=160pt,bblly=269pt,bburx=453pt,bbury=524pt,file=hawthorn2.ps,angle=0,height=70mm,width=70mm}}}
\caption{
(a) Short exposure [OIII]500.7nm emission-line halo seen above 
and below
the dustlane in Cen A, obtained with the `straddle shuffle' mode using
the TTF (see text). These are raw data: the bad column arises from trapped
charge during the charge shuffling. 
(b) Line $+$ continuum 
image of Cen A in order to emphasize the dust lane. 
}
\label{fig6}
\end{figure}

Extended ionization cones are seen in a variety of sources, including
the radio galaxies studied by Tadhunter (see above).  One of us has
obtained a short exposure [OIII] image of Cen A using the straddle
shuffle mode in order to achieve a clean subtraction of the elliptical
galaxy. In Fig.~6, we find a very highly ionized halo of
[OIII] emission above and below the dust lane. The component was first
discovered by Phillips\etal\ (1984) who found that the faint emission
appeared to rotate slowly compared to the dust lane, and typical line
dispersions are hundreds of \kms\ FWHM. We suspect that this emission
arises from a highly ionized wind which encompasses the radio jet over
a much larger solid angle.

Cecil\etal\ (2000) show that the famous braided jets in NGC 4258, which
can be traced over 10 kpc or more, emanate from the nuclear regions.
These jets have been seen in radio continuum, Rosat/Chandra x-rays, and
optical emission.  They obtained deep \Ha\ observations in the outer HI
disk which gave the first detailed picture of faint tendrils of the jets. 
A re-analysis of the radio continuum data has revealed
that these faint features have radio counterparts.

The NGC 1068 and NGC 4258 observations were able to detect faint
emission unseen previously with less sensitive imaging.  It is
important to realize that both observations were taken within 20\deg\
of the full moon, {\it which demonstrates the effectiveness of the TTF
in any phase of the lunar cycle.}

A number of powerful starburst galaxies show evidence for large-scale
winds along their minor axis (see reviews by Veilleux\etal\ 2001 and Heckman
2002). This phenomenon appears to be common at low and high redshift
(Lehnert \& Heckman 1996; Veilleux\etal\ 2001; Pettini\etal\ 2001).  The
most detailed studies have concentrated on objects like M82, NGC 253
and NGC 3079. It remains unclear just how much energy, mass and metals
these objects contribute to the intergalactic medium (IGM).  In some
cases, it is not clear if the wind is driven by a central starburst, an
AGN or a combination of these. However there is mounting evidence that the
optical diagnostics may greatly underestimate the true wind energetics
(Strickland \& Stevens 2000).

Veilleux \& Rupke (2002) provide spectacular evidence for a large-scale
wind in the edge on, early type galaxy NGC 1482. They obtain a clear
separation of disk material from the outflowing gas. The kinematics
show all the hallmarks of a biconal outflow. More impressively, the
entire emission line complex has diagnostics which are entirely
consistent with fast shocks. In other wind systems, there is often a
large contribution from the central ionizing stellar radiation field.

More recently, Veilleux\etal\ (2003) have presented results of a pilot
imaging study of 10 nearby starburst and active galaxies conducted with
the TTF on the Anglo-Australian Telescope.  An important goal of this
study was to search for warm emission-line gas on scales larger than
$\sim$ 10 kpc to constrain the zone of influence of these galaxies.
Large-scale structures are discovered or confirmed in four of the
galaxies (NGC~1068, NGC~1482, NGC~4388, NGC~6240, and NGC~7213).
Unsuspected structures are also seen for the first time in the galactic
winds of NGC~1365, NGC~1705, Circinus, and ESO484-G036. The TTF data
are combined with new optical long-slit spectra as well as published
and archived radio and X-ray maps to constrain the origin and source of
ionization of these filaments. A broad range of phenomena is observed,
including large-scale ionization cones and galactic winds, tidal
interaction, and ram-pressure stripping by an intracluster medium.

\subsection{Disk-halo connection in spirals}

For his thesis work, Miller (2002) has undertaken an emission line
survey of edge-on spiral galaxies in order to trace the connection
of disk star formation with activity in galactic haloes. Miller
\& Veilleux (1999) and Veilleux\etal\ (2001) show spectacular examples 
of diffuse ionized gas several kiloparsecs off the plane of
normal spirals. The ratio maps show enhanced [NII] emission with
respect to \Ha\ as we move away from the plane which is most easily
explained as a higher electron temperature in the halo gas (cf.
Sokolowski 1993; Reynolds\etal\ 1999; Collins \& Rand 2001). The source
of the temperature increase is a topic of debate in contemporary
astrophysics. Miller \& Veilleux (2003$a,b$) confirm a strong correlation
between the mass of ionized halo gas and the star formation rate per
unit disk area.
\begin{center}
\begin{figure}
\psfig{bbllx=-40pt,bblly=190pt,bburx=486pt,bbury=711pt,file=ferguson.ps,angle=0,height=100mm,width=120mm}\caption{
R band continuum images of elliptical/S0 galaxies (left) and their 
associated \Ha\ emission line images (right) from the TTF survey
of Ferguson\etal\ (2001).
}
\label{fig7}
\end{figure}
\end{center}
\subsection{Star formation in elliptical and spiral galaxies}

Very little is known about the frequency and nature of star formation
in elliptical galaxies, even though we have known for a 
long time that a large fraction contain cold gas (van Gorkom 1997;
Knapp 1999). Ferguson and collaborators (2001) have been making \Ha\
observations with TTF of a large sample of HI-selected ellipticals.
Essentially all of these systems (see Fig.~7) show evidence of star
formation to date.

For her thesis work, Cianci (2003$a,b$) is undertaking  a detailed
comparison of 20 face-on spirals observed at \Ha\ and \Hb, and UV
(150nm) images from the Ultraviolet Imaging Telescope (see Fig.~8). Her
particular interest is to understand asymmetries in spiral galaxies,
and the connection of the diffuse ionized gas with the HII regions and
dust distribution (cf. Zurita 2001; Zurita\etal\ 2000).

While spiral galaxies are commonly assumed to be axisymmetric, it is
surprising how few galaxies approach this ideal. The origin of 
asymmetries may have profound implications for their properties.
An interesting possibility is that the spiral galaxy disk `sloshes
around' within the dark matter potential (e.g. Syer \& Tremaine
1996; Schoenmakers, Franx \& de Zeeuw 1997) or that the dark halo
is lopsided (Jog 1999). Simulations show that the effects can
survive for one or two rotation periods ($\sim 10^8$ yr).

To date, studies of lopsidedness have concentrated on gas morphology
and dynamics. But if the simulations are right, asymmetric variations 
should show up in the young stellar population of any lopsided spiral.
Cianci\etal\ (2003) may now have detected an $m=1$ asymmetry in the 
spiral arms of the interacting galaxies M83 and M51. The UV continuum is
enhanced relative to H$\alpha$ along spurs that fall on the outside
of the western arm. Simulations with Starburst99 indicate that these
regions are older than 20 Myr, compared to the H-alpha bright regions
which appear to be younger than 7 Myr.

NGC 2915 is a dark-matter dominated, blue compact dwarf galaxy with an
HI disk extending to a radius of 15 kpc. In optical continuum,
it is a relatively nondescript with an optical radius of only 3 kpc. 
No stars were known to exist beyond this radius. Meurer\etal\ (1999)
obtained deep \Ha\ imaging with the TTF revealing, for the first 
time, faint HII regions at projected radii of 3 to 6 kpc. These have
since been confirmed with the Advanced Camera for Surveys on the 
Hubble Space Telescope (Meurer\etal\ 2003).

Higdon\etal\ (1997) have undertaken a multi-line study of the
Cartwheel galaxy. This is the most spectacular of the class of `ring
galaxies' which are excited by the central impact of an interloper. They
identify up to 100 star forming regions throughout the disk which they
will model as propagating star formation. There is a related class of
galaxies excited by off-centre impacts. One of these, NGC 1512, has
been looked at in detail with the TTF by F. Briggs. Moreover, there is
a wider class of objects under the general heading `event driven star 
formation'  (strong mergers, jet induced star formation, etc.) which
are relatively unexplored with the TTF.

\begin{figure}
\vskip 2.7in
\includegraphics{cianci.ps}
\caption{
A demonstration of the superb contrast possible when observing
a face-on spiral through a narrow band tuned to \Ha\ (left) compared to
a neighbouring continuum band (right). Both images were taken with the 
TTF and come from Cianci's (2002) spiral galaxy survey.
}
\label{fig8}
\end{figure}

\subsection{Stellar populations in galaxies}

One of the most important areas where very little work has been
done is imaging in stellar absorption lines. Beauchamp \& Hardy
(1997) have demonstrated that this is possible even with small
aperture telescopes. Molla, Hardy \& Beauchamp (1999) emphasize the
importance of spatially resolved, stellar absorption line mapping
of face-on spiral galaxies. 

Ryder, Fenner \& Gibson (2003) have now mapped two spirals, NGC 6221 and
NGC 7213, using the
Mg\_2 $\lambda5176$ and Fe $\lambda5270$ features, two
of the most prominent Lick/IDS indices (Worthey et al. 1994). The Lick
spectral indices do not measure abundances per se, but can be
transformed to quantities like [Fe/H] via the method of spectral
synthesis. The late-type spiral NGC 6221 shows a steep increase in
both indices at the core, before dropping back to the nuclear values
further out. This is in contrast to the early-type spiral NGC 7213
which has relatively constant indices in the core, followed by a
smooth rise in the outer disk.

Bland-Hawthorn\etal\ (2001) provide a list of stellar absorption 
line projects which should be considered for tunable filter work.

\section{GALACTIC SOURCES}

\subsection{Pulsar wind nebulae}

Jones\etal\ (2002$a$) discovered an extended \Ha\ nebula associated
with a pulsar moving rapidly through the interstellar medium. 
Several of these sources are now known, but in the case of 
pulsar B0740-28, the conic nebula is pinched perpendicular to
the long axis. Possible interpretations are variations in the
pulsar wind or in the external ISM.

\begin{figure}
\vskip 2.7in
\includegraphics{burton1.ps}
\includegraphics{burton2.ps}
\caption{
Schuberth \& Burton (2000) note that the left image is the first true map 
of carbon emission ever obtained. It was made possible by the TTF observing
in an ultranarrow band set to transmit the [CI]872.7nm emission line. 
Note the remarkable similarity with the Extremely Red Emission (ERE)
shown on the right, although the ERE emission is generally brighter 
further out from the central source.
}
\label{fig9}
\end{figure}

\subsection{Interstellar medium}

Schuberth \& Burton (2000) used the TTF to observe arguably the
most exotic line to date in the optical, the [CI]872.7nm line
buried deep within the OH forest. This line is extremely important
in photodissociation regions (PDR). NGC 2023 is an HII region on the 
surface of a molecular cloud. These authors postulated that
carbon atoms emitting at 872.7nm might delineate the PDR. Their
[CI] image (see Fig.~9) is the first true image of extended
carbon emission ever obtained. They show that the emission correlates
very closely to the so-called mysterious Extremely Red Emission (ERE)
at about 650nm. [CI] is clearly a powerful tool for probing galactic
star forming regions.

\subsection{Weather in brown dwarfs}

There is presently a lot of interest in detecting variability from
brown dwarfs. If the atmospheres and environs are cool enough, models
suggest that there should be dust condensations which swirl around
the brown dwarf producing variability in the light curve.  This was
detected for the first time by Tinney \& Tolley (1999) in the 
M-type brown dwarf LP 944-20. They used TTF in a novel set up which
required the charge to step on the CCD in synchrony with the TTF being
tuned between different bands.  Since then, variability has been found
in another brown dwarf Kelu-1 (Clarke 2002; Clarke, Tinney \& Covey 2002).
These sources do indeed appear to have complex weather patterns.

\subsection{Variable stars}

Deutsch\etal\ (1998) have used TTF in time series mode to
study the [OI]844.6nm emission line from the x-ray binary star
V2116 Oph, the optical counterpart of GC1+4. The symbiotic-like
optical spectrum of V2116 Oph shows the presence of a red giant.
The x-ray source is highly variable, so the [OI] line was studied
in the hope of detecting variability. This was rejected to a high
level of confidence. It is thought that the system is being seen
at a special time, and is probably an x-ray pulsar undergoing 
rapid evolution.

\subsection{Planets}

Ryder (2001) was able to image Mars in four narrow bands, a difficult
experiment since the source is so bright. With four bands set at 390,
500, 668 and 707nm, he produced high quality images of the ice caps,
the maria and terae, and various cloud formations. Jeremy Bailey (2003,
private communication) has recently emphasized the importance of
tunable filters in monitoring weather patterns on Mars and Venus in
narrow spectral bands. He compares the relative merits of a scanning
slit, a tunable filter and an integral field spectrograph on an 8$-$10m
telescope. The demands on planetary science are (i) spatial stability,
(ii) spatial integrity, (iii) spatial resolution of 25~mas or better, 
(iv) ability to switch rapidly between bands to maximize the time
resolution, and to combat the planet's rotation. All of
these requirements are met with a near-infrared tunable filter
(resolving power $R=200-1000$) used in conjunction with adaptive
optics on a large telescope.

\section{FUTURE PROSPECTS}

There are numerous ways in which tunable filters can find new and
important uses in astronomical programs.  One possible advance for tunable
filters are devices which can operate at cryogenic temperatures. Problems
like piezo creep and slow gain at cryo temperatures can now be overcome.

The largest tunable filter that Queensgate can make (150mm aperture)
is not well suited to 8--10m telescopes. But there is no reason why a
300mm aperture etalon, for example, cannot be made. We have an initial
design for such a system that uses an additional control stack at the
centre of the plates, well matched to the central obstruction of most
telescopes. This design requires a slightly reconfigured CS100 control
system.

It is commonly thought that tunable filters are restricted to 
switching between bands which fall within the restricted bandpass of
the order sorting filter. In fact, a multiband order sorter (Offer
\& Bland-Hawthorn 1998) allows spectral bands at opposite ends of the
optical spectrum to be observed in sequence (Cianci\etal\ 2000).

There exists a wide class of sources which have never benefitted from
detailed narrowband imaging largely because even through conventional
narrow bands, the field stars saturate the detector. One of many examples
was an attempt to measure gas metallicities in the vicinity of the
Trapezium. But rapid switching through differential ultranarrow bands
makes this entirely feasible now. The straddle shuffle demonstration
by Maloney \& Bland-Hawthorn (2001) has been used on galaxies with very
bright disks and cores, leading to perfect cancellation of the continuum
light and therefore revealing very weak levels of line emission. This
same method should work well on young globular clusters in order to
search for \Ha\ emission, but this has not been attempted to date.

Tunable filters have major applications for adaptive optics imaging
and for coronographic imaging. The use of differential narrowband
imaging would greatly help to suppress any stray light around the central
or nearby bright stars. B. Woodgate has achieved some spectacular
observations of emission line nebulae around symbiotic stars with
the Hubble Space Telescope, but the central stars are so bright that
diffraction spikes are a major problem. 

There are numerous other applications, some of which require minor
modifications to the existing hardware. These include new time 
series modes, nod \& shuffle imaging, mask \& shuffle imaging,
broad-narrow switching, tunable polarimetric imaging (all discussed in Bland-Hawthorn
2000) and tunable echelle imaging (Baldry \& Bland-Hawthorn 2000).

\medskip
\medskip
\appendix
\section {OBSERVING MODES OF TAURUS}

There are web sites which describe the basic observing modes
of Taurus:

\bigskip
\noindent
{\bf Taurus homepage:} {\tt www.aao.gov.au/taurus}

\noindent
{\bf Taurus Tunable Filter:} {\tt www.aao.gov.au/ttf}

\noindent
{\bf Taurus++:} {\tt www.aao.gov.au/taurus++}

\noindent
{\bf Taurus polarimetry:} {\tt www.aao.gov.au/taurus\_pol}

\section {TECHNICAL PAPERS RELATING TO THE TTF}

\noindent
{\bf General review on tunable filters:}
Bland-Hawthorn (2000$a,b$), Bland-Hawthorn\etal\ (2001)

\noindent
{\bf TTF instrument summary:} 
Bland-Hawthorn \& Jones (1998; 1999),
Jones \& Bland-Hawthorn (1998), Jones (2001$a$).

\noindent
{\bf Charge shuffle modes:}
Bland-Hawthorn \& Barton (1995),
Glazebrook \& Bland-Hawthorn (2001),
Maloney \& Bland-Hawthorn (2001), Jones (2001$b$).

\noindent
{\bf TTF data analysis:}
Jones\etal\ (2002$b$).

\section*{Acknowledgments}

\section*{References}

\reference Alexander, T.\etal\ 2000, ApJ, 536, 710
\reference Baker, J.C.\etal\ 2001, AJ, 121, 1821
\reference Baldry, I.K., Bland-Hawthorn, J. 2000, PASP, 112, 1112
\reference Beauchamp, D. \& Hardy, E. 1997, AJ, 113, 1666
\reference Birkinshaw, M., Worrall, D.M., Hardcastle, M.J. 2001. In New Visions of the X-ray Universe, ESTEC, in press
\reference Bland-Hawthorn, J., Barton, J. 1995, AAO Newsletter, 75, 8
\reference Bland-Hawthorn, J., Jones, D. H. 1998, PASA, 15, 44
\reference Bland-Hawthorn, J., Jones, D. H. 1998, SPIE, 3355, 855
\reference Bland-Hawthorn, J. 2000$a$, In Encyclopaedia of Astronomy \& Astrophysics, (IOP Publ. \& Macmillan Publ.: Bristol), ed. P.G. Murdin (see http://nedwww.ipac.caltech.edu/level5/Hawthorn2/frames.html)
\reference Bland-Hawthorn, J. 2000$b$, In Imaging the Universe in Three Dimensions, eds. W. van Breugel, J. Bland-Hawthorn, ASP Conf Ser, 195, 34
\reference Bland-Hawthorn, J.\etal\ 2001, ApJ, 563, 611
\reference Bowyer, S., Bergh\"{o}fer, T.W., Korpela, E.J. 1999, ApJ, 526, 592
\reference Cecil, G. 2000, SPIE, 4008, 73
\reference Cecil, G.\etal\ 2000, ApJ, 536, 675
\reference Cecil, G.\etal\ 2001, BAAS, 199, 5011
\reference Cecil, G.\etal\ 2002, ApJ, 568, 627
\reference Cepa, J.\etal\ 2002, In Galaxies: the Third Dimension, eds M Rosada, L Binette, L Arias, in press
\reference Cianci, S. 2000, AAO Newsletter, 94, 3
\reference Cianci, S.\etal\ 2000, In Imaging the Universe in Three Dimensions, eds. W. van Breugel, J. Bland-Hawthorn, ASP Conf Ser, 195, 601
\reference Cianci, S. 2003$a$, AAO Newsletter, 102, 4
\reference Cianci, S. 2003$b$, PhD, Univ. of Sydney
\reference Cianci, S. \etal\ 2003, ApJ, submitted
\reference Clarke, F. 2002, PhD, Univ. of Cambridge
\reference Clarke, F., Tinney, C.G., Covey, K.R. 2002, MNRAS, in press (astro-ph/0201162)
\reference Collins, J., Rand, R. 2001, ApJ, 551, 57
\reference Couch, W.J., Newell, E.B. 1984, ApJS, 56, 143
\reference Couch, W.J.\etal\ 2001, ApJ, 549, 820
\reference Dalcanton, J.J. 1996, ApJ, 466, 92
\reference Deutsch, E.W.\etal\ 1998, PASP, 110, 912
\reference Dressler, A., Gunn, J.E. 1983, ApJ, 270, 7
\reference Ferguson, A., van der Hulst, T., van Gorkom, J. 2001, AAO Newsletter, 96,4
\reference Francis, P., Wilson, G.M. \& Woodgate, B.E. 2001$a$, PASA, 18, 64
\reference Francis, P.\etal\ 2001$b$, ApJ, 554, 1001
\reference Fujita, S.S.\etal\ 2003, ApJL, in press (astro-ph/0302473)
\reference Glazebrook, K. 1997, In Extragalactic Astronomy in the Infrared, Moriond, eds. GA Mamon, TX Thuan, p. 467
\reference Glazebrook, K., Bland-Hawthorn, J. 2001, PASP, 113, 197
\reference Heckman, T. 2002, In Extragalactic Gas at Low Redshift, eds JS Mulchaey and JT Stocke, ASP Conf Ser, 254, 292
\reference Hewett, P.C.\etal\ 2000, In Imaging the Universe in Three Dimensions, eds. W. van Breugel, J. Bland-Hawthorn, ASP Conf Ser, 195, 94
\reference Higdon, J.L, Cecil, G. \& Bland-Hawthorn, J. 1997, AAS, 191, 8907
\reference Hippelein, H.\etal\ 2003, A\&A, accepted (astro-ph/0302116)
\reference Hu, E.M., McMahon, R.G. 1996, Nature, 382, 231
\reference Jaffe, W., Bremer, M. 2000, AAO Newsletter, 93, 3
\reference Jog, C.J. 1999, ApJ, 522, 661
\reference Jones, D.H., Bland-Hawthorn, J. 1998, PASP, 110, 1059
\reference Jones, D.H., Bland-Hawthorn, J. 1999, In Looking Deep in the Southern Sky, eds R Morganti, WJ Couch, p. 320
\reference Jones, D.H. 2001, PhD, Australia National University
\reference Jones, D.H. 2001$a$, PASP, 113, 255
\reference Jones, D.H. 2001$b$, In Scientific Drivers for ESO Future VLT/VLTI Instrumentation, eds. J. Bergeron and G. Monnet
\reference Jones, D.H., Bland-Hawthorn, J. 2001, ApJ, 550, 593
\reference Jones, D.H., Renzini, A., Rosati, P., Seifert, W. 2001, ESO Mess, 103, 10
\reference Jones, D.H., Stappers, B. \& Gaensler, B. 2002$a$, A\&A, in press
\reference Jones, D.H.\etal\ 2002$b$, MNRAS, 329, 759
\reference Knapp, J. 1999, In Star Formation in Early-type Galaxies, ASP Conf. Ser., 163, 119
\reference Lehnert, M.D. \& Heckman, T. M. 1996, ApJ, 462, 651
\reference Lieu, R.\etal\ 1999, ApJ, 527, L27
\reference Maloney, P.R. \& Bland-Hawthorn, J. 2001, ApJ, 553, L129
\reference Meurer, G.\etal\ 1999, BAAS, 194, 501
\reference Meurer, G.\etal\ 2003, IAU Symp. 220, Dark Matter In Galaxies
\reference Miller, S.T., Veilleux, S. 1999, AAO Newsletter, 88, 4
\reference Miller, S.T. 2002, PhD, Univ. of Maryland
\reference Miller, S.T. \& Veilleux, S. 2003$a$, ApJS, in press (astro-ph/0305026)
\reference Miller, S.T. \& Veilleux, S. 2003$b$, ApJ, in press (astro-ph/0304471)
\reference Miralde-Escud\`{e}, J., Lehar, J. 1992, MNRAS, 259, 31P
\reference Molla, M., Hardy, E. \& Beauchamp, D. 1999, ApJ, 513, 695
\reference Offer, A., Bland-Hawthorn, J. 1998, MNRAS, 299, 176
\reference Pettini, M.\etal\ 2001, ApJ, 554, 981
\reference Phillips, M.M.\etal\ 1984, Nature, 310, 554
\reference Pier, E.A., Krolik, J.H. 1993, ApJ, 418, 673
\reference Pogge, R.W. 1988, ApJ, 328, 519
\reference Reynolds, R., Haffner, M., Tufte, S. 1999, ApJ, 525, L21
\reference Riddick, F., Lucas, P. \& Roche, P. 2003, AAO Newsletter, 102, 10
\reference Rupke, D.\etal\ 2002, in prep
\reference Ryder, S. 2001, AAO Newsletter, 98, 11
\reference Ryder, S., Fenner, Y. \& Gibson, B.K. 2003, AAO Newsletter, 102, 6
\reference Schoenmakers, A.P., Franx, M. \& de Zeeuw, P.T. 1997, MNRAS, 292, 349
\reference Schuberth, Y., Burton, M.G. 2000, AAO Newsletter, 92, 3
\reference Shopbell, P.L.\etal\ 1999, ApJ, 524, L83
\reference Shopbell, P.L.\etal\ 2001, BAAS, 198, 5706
\reference Smail, I.\etal\ 1998, MNRAS, 293, 124
\reference Sokolowski, J. 1993, PhD, Rice University
\reference Solorzano-Inarrea, C.\etal\ 2002, MNRAS, 331, 673
\reference Solorzano-Inarrea, C. \& Tadhunter, C.N. 2003, MNRAS, accepted (astro-ph/0301157)
\reference Strickland, D.K., Stevens, I. R. 2000, MNRAS, 314, 511
\reference Syer, D. \& Tremaine, S. 1996, MNRAS, 281, 925
\reference Tadhunter, C.N., Tsvetanov, Z. 1989, Nature, 341, 422
\reference Tadhunter, C.N.\etal\ 2000, MNRAS, 314, 849
\reference Thommes, E.\etal\ 1998, MNRAS, 293, L6
\reference Tinney, C.G., Tolley, A.J. 1999, MNRAS, 304, 119
\reference Tober, J., Glazebrook, K., Thomson S., Bland-Hawthorn, J. \& Abraham, R.G. 2003, ApJ, submitted
\reference Tresse, L., Maddox, S.J., Le Fevre, O., Cuby, J.-G. 2001, MNRAS, 337, 369
\reference van Gorkom, J. 1997, In The Nature of Elliptical Galaxies, ASP Conf Ser, 116, 363
\reference Veilleux, S., Bland-Hawthorn, J. 1997, ApJ, 479, L105
\reference Veilleux, S. 2002$a$, In Extragalactic Gas at Low Redshift, eds JS Mulchaey and JT Stocke, ASP Conf Ser, 254, 313
\reference Veilleux, S. 2002$b$, In Galaxies: the Third Dimension, eds M Rosada, L Binette, L Arias, in press
\reference Veilleux, S., Rupke, D. 2002, ApJ, 565, L63
\reference Veilleux, S., Cecil, G., Bland-Hawthorn, J. \& Shopbell, P.L. 2002, In Line Emission from Jet Flows, eds. W. Henney, W. Steffen, L. Binette \& A. Raga (Isla Mujeres), {\it Rev. Mex. Astron. Astrophys.}, 13, 222--229
\reference Veilleux, S., Shopbell, P.L., Rupke, D. \& Bland-Hawthorn, J. 2003, ApJ, submitted
\reference Veilleux, S. \& Miller, S.T. 2003, AAO Newsletter, 102, 8
\reference Young, A.J., Wilson, A.S., Shopbell, P.L. 2001, ApJ, 556, 6
\reference Zaritsky, D.\etal\ 1997, ApJ, 480, L91
\reference Zurita, A. 2001, PhD thesis, Instituto Astrofisica, La Laguna
\reference Zurita, A.\etal\ 2000, A\&A, 363, 9

\end{document}